# Characterization of the errors of the Fast Multipole Method approximation in particle simulations

Felipe A. Cruz [*]

Department of Mathematics

University of Bristol

Bristol, United Kingdom BS8 1TW

L. A. Barba [†]

Department of Mathematics

University of Bristol

Bristol, United Kingdom BS8 1TW

1 April, 2008

## Abstract

The Fast Multipole Method (FMM) offers an acceleration for pairwise interaction calculation, known as $N$-body problems, from $\mathcal{O}(N^2)$ to $\mathcal{O}(N)$ with $N$ particles. This has brought dramatic increase in the capability of particle simulations in many application areas, such as electrostatics, particle formulations of fluid mechanics, and others. Although the literature on the subject provides theoretical error bounds for the FMM approximation, there are not many reports of the measured errors in a suite of computational experiments. We have performed such an experimental investigation, and summarized the results of about 1000 calculations using the FMM algorithm, to characterize the accuracy of the method in relation with the different parameters available to the user. In addition to the more standard diagnostic of the maximum error, we supply illustrations of the spatial distribution of the errors, which offers visual evidence of all the contributing factors to the overall approximation accuracy: multipole expansion, local expansion, hierarchical spatial decomposition (interaction lists, local domain, far domain). This presentation is a contribution to any researcher wishing to incorporate the FMM acceleration to their application code, as it aids in understanding where accuracy is gained or compromised.

---

[*]Acknowledges support from the SCAT project via EuropeAid contract II-0537-FC-FA, www.scat-alfa.eu

[†]Acknowledges support from EPSRC under grant contract EP/E033083/1.

# 1   Introduction

In particle simulations, often the dynamics results from the summation of all pair-wise forces in the ensemble of particles. Such situations arise in astrophysics, molecular dynamics, plasma physics, and certain formulations of fluid dynamics problems, for example. The total field of interest (gravitational, electrostatic, etc.) at one evaluation point requires adding the contribution of all source points or particles, and so if both evaluation points and particles are numbered at $N$, a total of $N^2$ operations is needed. This fact was for a long time an impediment to the wider use of particle simulations, as the computational effort becomes prohibitive for large numbers of particles.

The above scenario changed dramatically with the introduction of tree-codes and the fast multipole method (FMM), which appear in the late 1980s for accelerated evaluation of $N$-body problems. Tree-codes [1, 4] are generally perceived to be easier to grasp and program, and provide a complexity of $\mathcal{O}(N \log N)$. The FMM was introduced as an algorithm for the rapid evaluation of gravitational or Coulombic interactions [11] and promises a reduction in computational complexity to $\mathcal{O}(N)$. It has, since its dissemination, been adapted for many applications: for fast evaluation of boundary elements [10], for vortex sheet problems with desingularized equations [12], for long-range electrostatics in DNA simulations [9], and many others. The impact of the FMM has been undeniably huge, resulting in it being chosen as one of the Top 10 Algorithms of the 20th Century [8].

Despite the great volume of work using and adapting the FMM in many application areas, there remains some lack of insight regarding how the algorithm can be efficiently used to obtain an accurate representation of the field of interest. The error of the FMM approximation is estimated by theoretical bounds, which as could be expected reveal a trade-off between accuracy and efficiency of the computation. However, there is not much literature providing measurements of the accuracy of the approximation, in practice. One may often find such assertions in published works as "only the first three moments of the expansion were used", or something to that effect. But just as often there is no information provided about the actual errors which are *observed*. Of course, it is not easy to provide such measures of observed error, as this would require additional computations using the direct $\mathcal{O}(N^2)$ method, for comparison purposes. Nevertheless, it is important for users of the algorithm to know what to expect in terms of accuracy and efficiency, depending on the choice of algorithm parameters.

We aim to contribute to this gap in understanding by presenting a methodical investigation into the errors of the approximation used by the FMM, when the underlying 'client' application is the calculation of the velocity field induced by $N$ regularized particles of vorticity. This application is rather more demanding than the Newtonian force calculation, because in the latter case the gravitational interaction is dominated by the first moment —due to the fact that all mass is positive. Therefore, keeping only the first three moments could easily give the desired accuracy. On the other hand, as in Coulomb electrostatic calculations, the vortex particles can be both positive and negative, and thus an acceptable accuracy may require that more terms in the expansion be kept.

For the purposes of this study, a prototype code of the FMM computation of the velocity induced by $N$ vortex particles was implemented using the Python[1] language. The nice features of Python —such as dynamic typing, extensive numerical libraries, and high programmer productivity— helped us produce a piece of software which is easy to use and easy to understand. We are currently using this Python code as a starting point for a parallel version of the code which, in collaboration with members of the developer team, will be incorporated to the PETSc library for scientific computing[2]. This project will be reported elsewhere, but preliminary results are being presented in the upcoming Parallel CFD meeting[7]. Our final aim is to contribute to the community of particle simulations with an open source FMM implementation which is parallel and portable. For the time being, the Python code is being made available publicly and we welcome correspondence from interested researchers[2].

---

[1] http://www.python.org/

Using the Python FMM code, more than 900 calculations were performed, varying the numerical parameters: $N$, the number of particles, $l$, the number of levels in the tree, and $p$, the truncation level of the multipole expansion. We looked not only at the maximum error in the domain, which would be the conventional approach; we also present results showing how the error varies in space, revealing some interesting features of the method. Through this presentation of the results, we believe a clear characterization of the nature of the FMM approximation is obtained.

The paper is organized as follows. The next section presents an outline of the vortex particle method, for completeness. In §3, we offer an overview of the FMM, with some details of our implementation. Following, in §4, we discuss briefly the sources of errors in the FMM algorithm. And finally, §5 reports the detailed experiments using the FMM for evaluation of the velocity of $N$ vortex particles; the behavior of the method will be illustrated for varying parameters, as well as the impact on the efficiency of the calculation, for different problem sizes.

## 2 Review of the vortex particle method

For incompressible flow one has that $\nabla \cdot \mathbf{u}(\mathbf{x}, t) = 0$, and the two-dimensional vorticity transport equation is expressed as,

$$\frac{\partial \omega}{\partial t} + \mathbf{u} \cdot \nabla \omega = \nu \Delta \omega. \tag{1}$$

The vortex particle method proceeds by spatially discretizing the vorticity field onto particles which have a radially symmetric distribution of vorticity, and thus the vorticity field is effectively approximated by a radial basis function expansion as follows:

$$\omega(\mathbf{x}, t) \approx \omega_\sigma(\mathbf{x}, t) = \sum_{i=1}^{N} \Gamma_i(t) \zeta_{\sigma_i} \left( \mathbf{x} - \mathbf{x}_i(t) \right). \tag{2}$$

Here, the $\mathbf{x}_i$ represent the particle positions, $\Gamma_i$ are the circulation values assigned to each vortex particle, and the core size is $\sigma_i$. The core sizes are usually uniform ($\sigma_i = \sigma$), and the characteristic distribution of vorticity $\zeta_{\sigma_i}$, commonly called the cut-off function, is frequently a Gaussian distribution, such as:

$$\zeta_\sigma(\mathbf{x}) = \frac{1}{2\pi\sigma^2} \exp\left(\frac{-|\mathbf{x}|^2}{2\sigma^2}\right). \tag{3}$$

The velocity field is obtained from the vorticity by means of the Biot-Savart law of vorticity dynamics:

$$\mathbf{u}(\mathbf{x}, t) = \int (\nabla \times \mathbf{G})(\mathbf{x} - \mathbf{x}')\omega(\mathbf{x}', t)d\mathbf{x}' = \int \mathbf{K}(\mathbf{x} - \mathbf{x}')\omega(\mathbf{x}', t)d\mathbf{x}' = (\mathbf{K} * \omega)(\mathbf{x}, t)$$

---

[2] http://www.maths.bris.ac.uk/~aelab/research/pyFMM.html

where $\mathbf{K} = \nabla \times \mathbf{G}$ is the Biot-Savart kernel, with $\mathbf{G}$ the Green's function for the Poisson equation, and $*$ representing convolution. For example, in 2D the Biot-Savart law is written explicitly as,

$$\mathbf{u}(\mathbf{x},t) = \frac{-1}{2\pi} \int \frac{(\mathbf{x}-\mathbf{x}') \times \omega(\mathbf{x}',t)\hat{\mathbf{k}}}{|\mathbf{x}-\mathbf{x}'|^2} d\mathbf{x}'. \tag{4}$$

The vorticity transport is solved in this discretized form by convecting the particles with the local fluid velocity, and accounting for viscous effects by changing the particle vorticity. Hence, the unbounded vortex method is expressed by the following system of equations:

$$\frac{d\mathbf{x}_i}{dt} = \mathbf{u}(\mathbf{x}_i,t) = (\mathbf{K} * \omega)(\mathbf{x}_i,t), \qquad \frac{d\omega}{dt} = \nu \nabla^2 \omega. \tag{5}$$

There are a variety of ways to account for the viscous effects represented by the second of these equations, of which a discussion is offered in [3]. See also the book [6]. But the convection of vorticity is always treated in the same way: by integration of the ordinary differential equations for the particle trajectories. In this process, it is necessary at each step to calculate the velocity at each particle location, using the Biot Savart law. Using a radially symmetric distribution function, one can obtain an analytic result of the integral that appears in (4), which results in an expression for the velocity at each point as a sum over all particles. For the 2D Gaussian particle distribution, one has:

$$\mathbf{K}_\sigma(\mathbf{x}) = \frac{1}{2\pi r^2}(-y,x)\left(1 - \exp(-\frac{r^2}{2\sigma^2})\right). \tag{6}$$

where $r^2 = x^2 + y^2$. Thus, the formula for the discrete Biot-Savart law in two dimensions gives the velocity as follows,

$$\mathbf{u}_\sigma(\mathbf{x},t) = \sum_{j=1}^{N} \Gamma_j \, \mathbf{K}_\sigma(\mathbf{x}-\mathbf{x}_j). \tag{7}$$

As can be seen in Equation (7), calculating the velocity at one point takes $N$ operations. Thus, calculating the velocity at every point is an $N$-body problem.

The fact that the direct evaluation of the velocity field induced by $N$ vortex particles is an $\mathcal{O}(N^2)$ calculation was for a long time an obstacle for wider acceptance of vortex methods. The dissemination of the fast multipole method quickly provided an opportunity to make vortex particle methods more efficient and perhaps even competitive with mainstream methods for direct numerical simulation. Indeed, the vortex particle method has matured in the last, say, 15 years, and we now have demonstrations of highly competitive applications, such as for example a recent calculation of aircraft trailing wakes involving a *billion* particles [5]. Such results would be impossible without the FMM approximation of the Biot-Savart calculation.

## 3 Implementation of the FMM

### 3.1 Overview of the algorithm

The fast multipole method (FMM), as well as its ancestor the treecode, is an algorithm for accelerating the computation of so-called $N$-body problems, that is, problems involving $N$ bodies or particles interacting with each other. Suppose that a function $f$ can be defined to describe the influence of all the particles of a system at any place; $f$ is written as the sum of pairwise interactions with a set of source particles, as follows:

$$f(y) = \sum_{i=1}^{N} c_i \mathbf{K}(y - x_i), \tag{8}$$

where $c_i$ are scalar values and $x_i$ represent the source particles. The assumptions when applying the FMM are:

1. The evaluation of $f(\cdot)$ occurs at a large number of *evaluation* points $\{y_i\}$. Typically, the evaluation points include the source points.

2. No self-interaction: a particle does not interact with itself.

3. It is common that the set of source points $\{x_i\}$ and the set of evaluation points $\{y_i\}$ contain about the same number of members.

4. The influence between two points, $\mathbf{K}(y - x)$, decays as the points are in some sense *far away*.

5. The kernel $\mathbf{K}$ that represents the pairwise interactions should be smooth far from the origin; such kernel, when evaluated far away from the source point, will enable us to compute the aggregate effect of a cluster of source points instead of computing all the pairwise interactions.

Under these conditions and evaluating at the source points, Equation (8) implies:

$$f(x_j) = \sum_{i=1, i \neq j}^{N} c_i \mathbf{K}(x_j - x_i) \tag{9}$$

With $N$ particles, the complexity of the direct calculation of Equation (9) at every evaluation point is $O(N^2)$.

Both the fast multipole algorithm and treecodes are based on the idea that the influence of a cluster of particles can be approximated by an agglomerated quantity, when such influence is evaluated far enough away from the cluster itself. The algorithms require a division of the computational domain into a *near-domain* and a *far-domain*:

**Near domain:** contains all the particles that are near the evaluation point, and is usually a minor fraction of all the $N$ particles. The influence of the near-domain is computed by directly evaluating the pair-wise particle interactions. The computational cost of directly evaluating the near domain is not dominant as the near-domain remains small.

**Far domain:** contains all the particles that are far away from the evaluation point, and ideally contains most of the $N$ particles of the domain. The evaluation of the far domain will be sped-up by evaluating the approximated influence of clusters of particles rather than computing the interaction with every particle of the system.

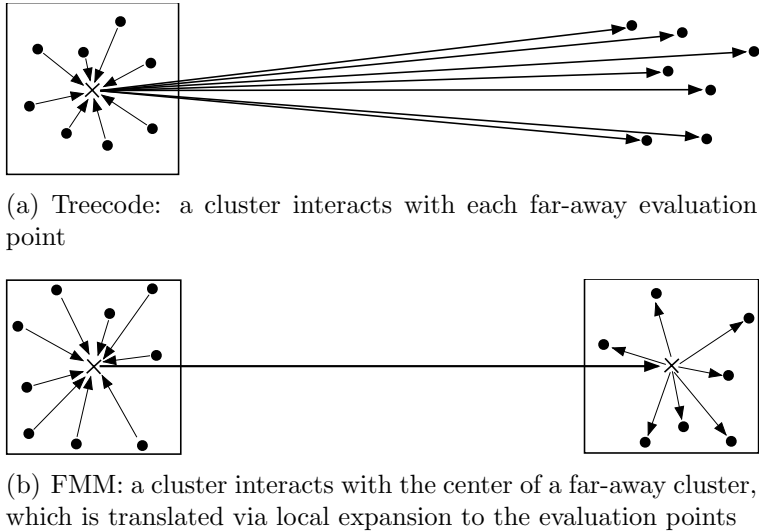

(a) Treecode: a cluster interacts with each far-away evaluation point

(b) FMM: a cluster interacts with the center of a far-away cluster, which is translated via local expansion to the evaluation points

Figure 1: Illustrations representing the treecode and FMM clustering techniques, resulting in the approximation of far-away particle influences.

The key idea behind the FMM is that particle influences can be clustered and expressed in two different representations: as Multipole Expansions (MEs) and as Local Expansions (LEs). The MEs and LEs are Taylor series that converge in different subdomains of space. The center of the series for an ME is the center of the cluster of source particles, and it only converges outside the cluster of particles. In the case of an LE, the series is centered near an evaluation point and converges locally.

In the case of treecodes, one does not have local expansions. Thus, in a treecode, a cluster of particles interacts with each far-away evaluation point, as illustrated in Figure 1(a). The difference with the FMM is illustrated in Figure 1(b), where a cluster's aggregated influence is locally translated to evaluation points, achieving a greater acceleration (in theory).

The first step of the FMM is to hierarchically subdivide space in order to form the clusters of particles; this is accomplished by using a tree structure, illustrated in Figure 2, to represent each subdivision. In a one-dimensional example: level 0 is the whole domain, which is split in two halves at level 1, and so on up to level $l$. In two dimensions, instead each domain is divided in four, to obtain a quadtree, while in three dimensions, domains are split in 8 to obtain an oct-tree. In all cases, we can make a flat drawing of the tree as in Fig. 2, with the only difference being the number of branches coming out of each node.

The next step is to build the MEs for each node of the tree; the MEs are built first at the deepest level, level $l$, and then translated to the center of the parent cell. This is referred to as the *upward sweep* of the tree. Then, in the *downward sweep* the MEs are first translated into LEs for all the boxes in the *interaction list*. At each level, the interaction list corresponds to the cells of the same level that are in the far field for a given cell. Finally, the LEs of upper levels are added up to obtain the complete far domain influence for each box at the leaf level of the tree. These ideas are better visualized with an illustration, as provided in Figure 3.

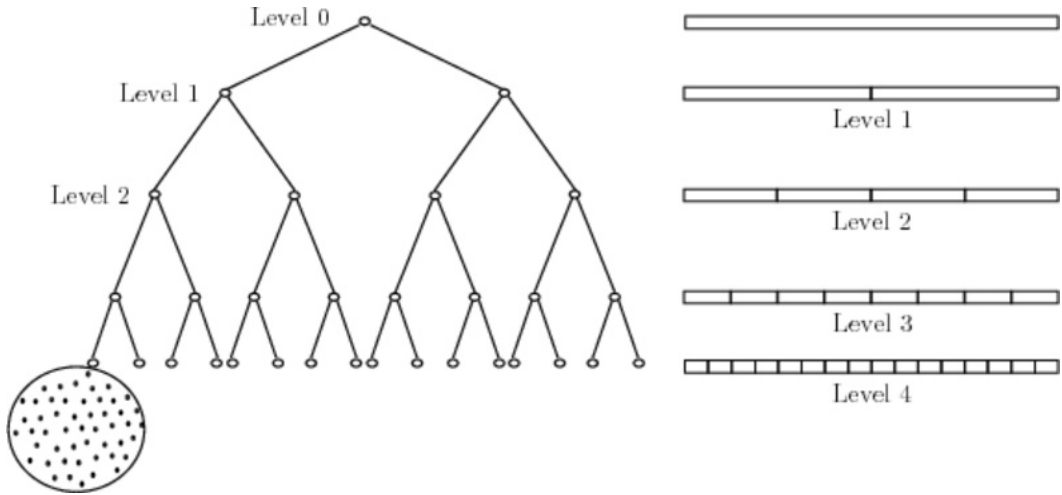

Figure 2: Sketch of a one-dimensional domain (right), divided hierarchically using a binary tree (left), to illustrate the meaning of levels in a tree and the idea of a final leaf holding a set of particles at the deepest level.

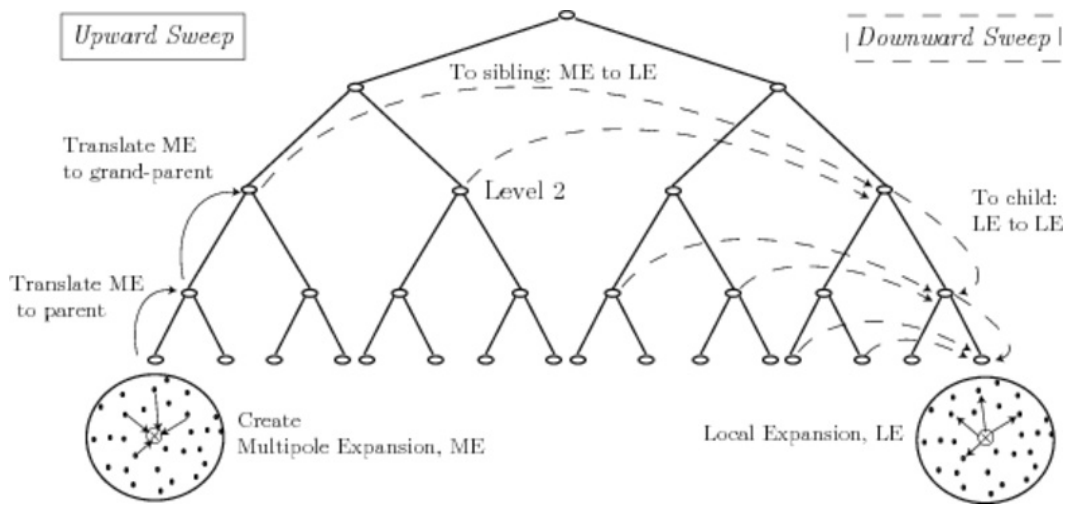

Figure 3: Illustration of the *upward sweep* and the *downward sweep* of the tree. The multipole expansions (ME) are created at the deepest level, then translated upwards to the center of the parent cells. The MEs are then translated to a local expansion (LE) for the siblings at all levels deeper than level 2, and then translated downward to children cells. Finally, the LEs are created at the deepest levels.

The total field at each evaluation point is thus the sum of the contributions of the near and far domain:

$$\begin{aligned} f(y) &= f^{\text{near}}(y) + f^{\text{far}}(y) \\ &= \sum_{x_i \text{ near } y} c_i \, \mathbf{K}(y - x_i) + \sum_{x_i \text{ far from } y} c_i \, \mathbf{K}(y - x_i) \end{aligned} \quad (10)$$

where the first summation of (10) performs the computation of the function $f$ using the direct method over the particles inside the near field. The second summation of (10) is approximated by the fast multipole method. To be able to use the FMM approximation, the kernel function $\mathbf{K}$ must allow the decoupling of the influence of the source particles $x_i$ and the evaluation points $y$, that is,

$$\mathbf{K}(y - x) = \sum_{m=0}^{\infty} a_m(x) f_m(y) \quad (11)$$

by replacing in (8), we obtain

$$\begin{aligned} f(y) &= \sum_{i=1}^{N} c_i \mathbf{K}(y - x_i) \\ &= \sum_{i=1}^{N} c_i \sum_{m=0}^{\infty} a_m(x_i) f_m(y) \\ &= \sum_{m=0}^{\infty} \left( \sum_{i=1}^{N} c_i a_m(x_i) \right) f_m(y) \end{aligned}$$

The series expansion of the kernel in (11) might not exactly represent or approximate $\mathbf{K}$ for arbitray values of $x$ and $y$, but we expect that for particles located in the far-field domain the effect of $\mathbf{K}$ will be smooth, and be approximately decoupled due to the fact that the source particles $\{x_i\}$ and the evaluation points are well separated. By truncating the infinite series in (11), the number of terms left after the truncation procedure will control the accuracy of the approximation:

$$f(y) = \sum_{m=0}^{p-1} \left( \sum_{i=1}^{N} c_i a_m(x_i) \right) f_m(y) + \mathbf{error}(p). \quad (12)$$

The terms that are dependent on the source points can be computed a single time at cost $\mathcal{O}(N)$. Then for $N$ evaluations of (12) only $\mathcal{O}(pN)$ operations are performed, instead of the original $\mathcal{O}(N^2)$ operations in (9). So far, we have introduced a general idea of how the FMM works, but much of the details of the final $\mathcal{O}(N)$ algorithm have been left out. For more details of the algorithm, we cite [11].

## 3.2 The FMM applied to the velocity field of $N$ vortex particles

### 3.2.1 Multipole Expansions

The expansion required by the Biot-Savart kernel, used to compute the velocity induced by $N$ vortex particles, is now explained. One of the first investigations of the approximation of this kernel by means of the FMM was presented in [12]. As in that work, we perform the computation of the velocity field in the complex plane $\mathbb{C}$. That is, given the position $z_i$ (complex variable) and circulation $\Gamma_i$ of a vortex particle, we compute the velocity it induces at each *evaluation* point $z$ using the Biot-Savart law, expressed as follows:

$$u_\sigma(z) = \sum_{i=1}^{N} \Gamma_i \, \mathbf{K}_\sigma(z - z_i) \tag{13}$$

$$\mathbf{K}_\sigma(z) = -\frac{1}{2\pi|z|^2} \begin{pmatrix} v \\ -u \end{pmatrix} \left(1 - \exp\left(\frac{-|z|^2}{2\sigma^2}\right)\right)$$

$$z = u + jv, \qquad u, v \in \mathbb{R}, z \in \mathbb{C}$$

If the evaluation point $z$ is located far away from the source particles $z_i$, this expression can be well approximated by

$$u(z) = \sum_{i=1}^{N} \Gamma_i \, \mathbf{K}(z - z_i) \tag{14}$$

$$\mathbf{K}(z) = -\frac{1}{2\pi|z|^2} \begin{pmatrix} v \\ -u \end{pmatrix} \tag{15}$$

$$z = u + jv, \qquad u, v \in \mathbb{R}, z \in \mathbb{C}$$

The kernel (15) can be written as

$$\begin{aligned}
\mathbf{K}(z) &= -\frac{1}{2\pi|z|^2} \begin{pmatrix} v \\ -u \end{pmatrix} \\
&= -\frac{1}{2\pi|z|^2}(v - ju) \\
&= -\frac{1}{2\pi j} \frac{(jv - j^2 u)}{|z|^2} \\
&= -\frac{1}{2\pi j} \frac{(u + jv)}{|z|^2} \\
&= \frac{j}{2\pi} \frac{z}{zz^*} \\
&= \frac{j}{2\pi z^*}
\end{aligned}$$

By replacing this result in (14) we obtain the complex conjugate velocity:

$$[u(z)]^* = \frac{j}{2\pi} \sum_{i=1}^{N} \frac{\Gamma_i}{z - z_i} \tag{16}$$

where $[\,\cdot\,]^*$ represents the conjugate operator of a complex number. So, computation of velocities when the evaluation point is located far away from the source point is just (16).

Next, we will describe the decoupling of the kernel in (16), where $x$ corresponds to a source particle and $y$ corresponds to an evaluation point. The decoupling aims to obtain an expression of the form:

$$\mathbf{K}(y - x) = \frac{1}{y - x} = \sum_{m=0}^{\infty} a_m(x) f_m(y) \tag{17}$$

$$x_i, y \in \mathbb{C}$$

In a similar way as in (12), we would like to find an expansion that for a given set of $\{x_i\}$, decouples the contributions of the $x_i$ around the center of the cluster $x_*$ and at the same time gives a good approximation of the kernel by using as few terms as possible. The following derivation accomplishes our aim:

$$\begin{aligned} \frac{1}{y-x} &= \frac{1}{(y-x_*)-(x-x_*)} \\ &= \frac{1}{(y-x_*)\left(1-\frac{x-x_*}{y-x_*}\right)} \\ &= \frac{1}{(y-x_*)} \sum_{m=0}^{\infty} \left(\frac{x-x_*}{y-x_*}\right)^m \end{aligned} \quad (18)$$

To obtain this result we used the following Taylor series expansion:

$$\frac{1}{1-a} = \sum_{m=0}^{\infty} a^m. \quad (19)$$

It is important to note that this sum only converges when $|a| < 1$ and diverges for any other values of $a$. For this reason, the expression obtained in (18) only converges when $|x - x_*| < |y - x_*|$. Now, we can rewrite (16) by using the result in (18):

$$\begin{aligned} [u(y)]^* &= -\frac{j}{2\pi} \sum_{i=1}^{N} \frac{\Gamma_i}{(y-x_*)} \sum_{m=0}^{\infty} \left(\frac{x_i - x_*}{y - x_*}\right)^m \\ &= \sum_{m=0}^{\infty} \left(\sum_{i=1}^{N} -\frac{j\Gamma_i}{2\pi}(x_i - x_*)^m\right) (y - x_*)^{-m-1} \end{aligned}$$

Finally,

$$[u(y)]^* = \sum_{m=0}^{\infty} \left(\sum_{i=1}^{N} c_i a_m(x_i, x_*)\right) f_m(y, x_*) \quad (20)$$

with

$$\begin{aligned} c_i &= -\frac{j\Gamma_i}{2\pi} \\ a_m(x_i, x_*) &= (x_i - x_*)^m \\ f_m(y, x_*) &= (y - x_*)^{-m-1} \end{aligned}$$

### 3.2.2 Truncating the Multipole Expansions

By truncating the series in (20), the ME becomes approximate, and the number of terms $p$ left after the truncation controls the accuracy of the approximation:

$$[u(y)]_p^* = \sum_{m=0}^{p-1} \left(\sum_{i=1}^{N} c_i a_m(x_i, x_*)\right) f_m(y, x_*) + \mathbf{error}(p; x_i, y) \quad (21)$$

Details on the error of truncation are presented in §4.

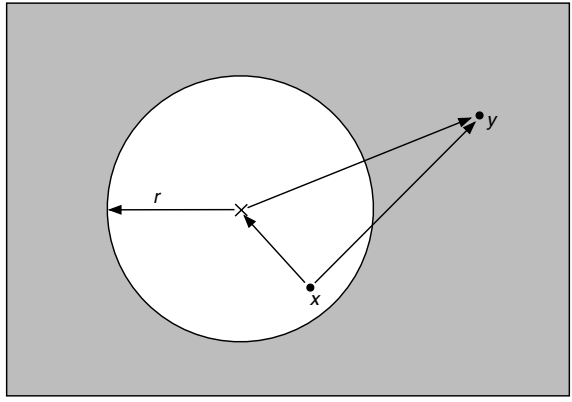

Figure 4: The ME is valid outside the circumference of center $x_*$ and radius $r$. An evaluation point $y$ lies outside, so every evaluation point must satisfy $|y - x_*| > r$.

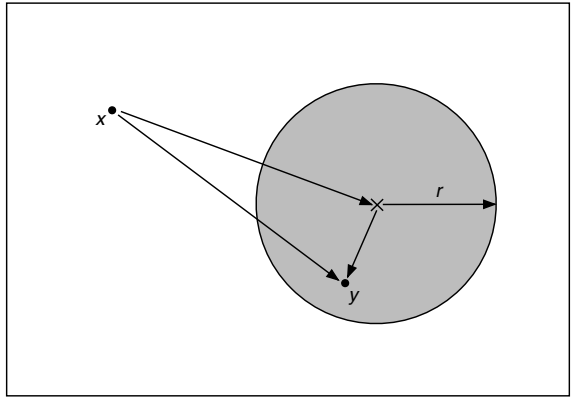

Figure 5: Multipole Expansion to Local Expansion.

### 3.2.3 Local Expansions

A tree-code [1, 4] is an algorithm that uses only Multipole Expansions to reduce complexity to $O(NlogN)$. In order to further reduce the complexity of the algorithm, a clever way to rapidly evaluate the Multipole Expansions is needed. The Local Expansion is now introduced.

From a practical point of view, the velocity induced by the particles located in the far field should be very similar for two evaluation points that are close enough. So, instead of singly evaluating the velocity at each point, we could reorganize the evaluation of the Multipole Expansions that build the far field into a *single expression* that represents the influence of the far field. This single expression is know as a Local Expansion (LE). The main characteristics of an LE are that it will represent the far field given by a set of MEs, and that an LE is only valid inside an evaluation domain.

An LE is a series expansion that is valid inside a circumference $C$, and provides a representation of the influence of a set of MEs that are centered far away from $C$. We need to explain now how we convert our ME into LE.

A Multipole Expansion has this formulation:
$$Me_k(y) = y^{-k-1}$$

A Local Expansion has this formulation:
$$Le_k(y) = y^k$$

The change of representation from a Multipole Expansion (ME) to a Local Expansion (LE) is obtained by rewriting the formulation of the ME into an LE:

$$\begin{aligned} Me_k(y+t) &= (y+t)^{-k-1} = t^{-k-1}\left(1+\frac{y}{t}\right)^{-k-1} \\ &= \sum_{n=0}^{\infty} \frac{(-1)^n(n+k)!}{n!k!} t^{-k-n-1} y^n \\ &= \sum_{n=0}^{\infty} \frac{(-1)^n(n+k)!}{n!k!} \frac{1}{t^{k+n+1}} Le_n(y) \end{aligned}$$

Rewriting the equation in a vector matrix representation, and truncating the series, we obtain:

$$\begin{aligned} Me(y+t) &= \sum_{n=0}^{p} (-1)^n \binom{n+k}{k} \frac{1}{t^{k+n+1}} Le_n(y) \\ &= M2L(t) Le(y) \end{aligned}$$

The expression $M2L(t)$ corresponds to a matrix that performs the conversion of the Multipole Expansion into a Local Expansion. The elements of the matrix are:

$$M2L(t)_{nk} = (-1)^n \binom{n+k}{k} \frac{1}{t^{k+n+1}} \tag{22}$$

## 3.3 Hierarchical space partitioning

Due to the spatial characteristics of the Multipole Expansion, we need a way to cluster the particles that are close in distance. The method that we will use to create the clusters is to construct a data structure that divides the domain into smaller sub-domains, where each sub-domain will be used to cluster the particles. For the two-dimensional case, the Fast Multipole Method uses a quad-tree, as represented on Figure 6, as the data structure for space partitioning and hierarchical ordering. One of the advantages of the quad-tree is that this hierarchical decomposition of the computational domain is simple to understand and, if implemented correctly, the quad-tree algorithm only takes $O(N)$ time to organize the $N$ particles. Even more, the use of this structure is convenient since the quad-tree nodes have a square shape that facilitates the Multipole Expansion creation and the calculation of the bounds for the errors. Also, for this type of structures it is easy to determine spatial relations between nodes that later will help us to determine the the near and far domains.

### 3.3.1 Spatial relations

The quad-tree data structure allows us to relate its nodes; the way in which the tree is constructed is especially useful to determine relations between the sizes of boxes and the distances between different levels. These relations are used by the FMM to determine the local and far domains relative to the set of particles in a box and to establish relations between the particles contained in different boxes (see Figures 7 and 8):

- *Parent node.* A box $b$ at level $l$ is *parent* of the boxes created by its direct subdivision and creation of boxes $\{b_0, b_1, b_2, b_3\}$ at level $(l+1)$. Each parent box has four children boxes.

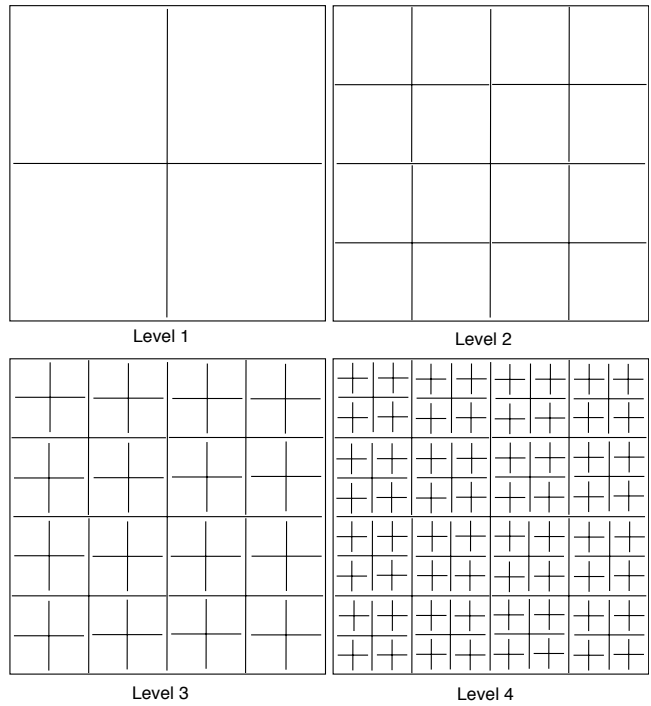

Figure 6: This figure shows the hierarchy of boxes created by the quad-tree. The top level, level 0, is an empty box that holds the complete computational domain. To obtain the next level, each box is subdivided into four new boxes, until the finest level is reached which is not subdivided. In this way, the four boxes seen at level 1 are the children of the single box at level 0; level 1 is subdivided to obtain the boxes of level 2, and so on for levels 3 and 4. A box $b$ at level $l$ will have four children at level $l+1$, box $b$ will be known as the parent box of its children. The number of distinct boxes at the quad-tree level $l$ is $4^l$.

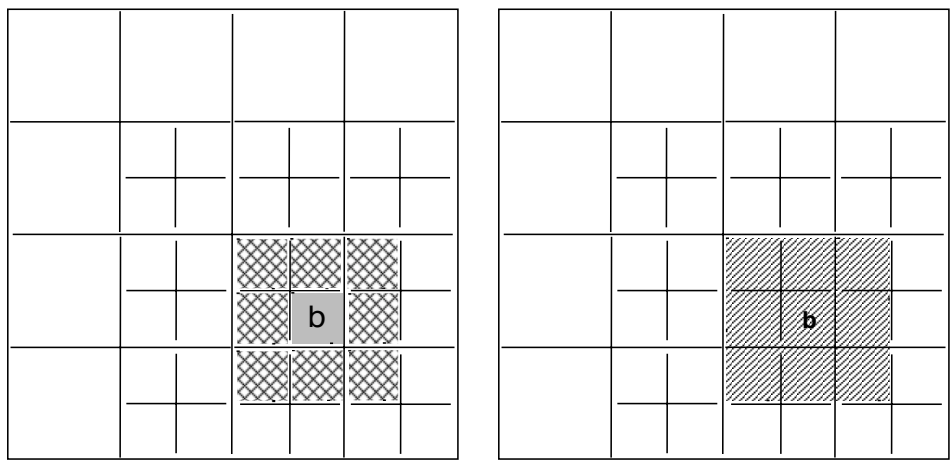

(a) Neighbors correspond to the boxes with the cross mesh design.

(b) Local list corresponds to the boxes with the dashed color.

Figure 7: Neighbors and local list of a box $b$ at level 3 in the quad-tree.

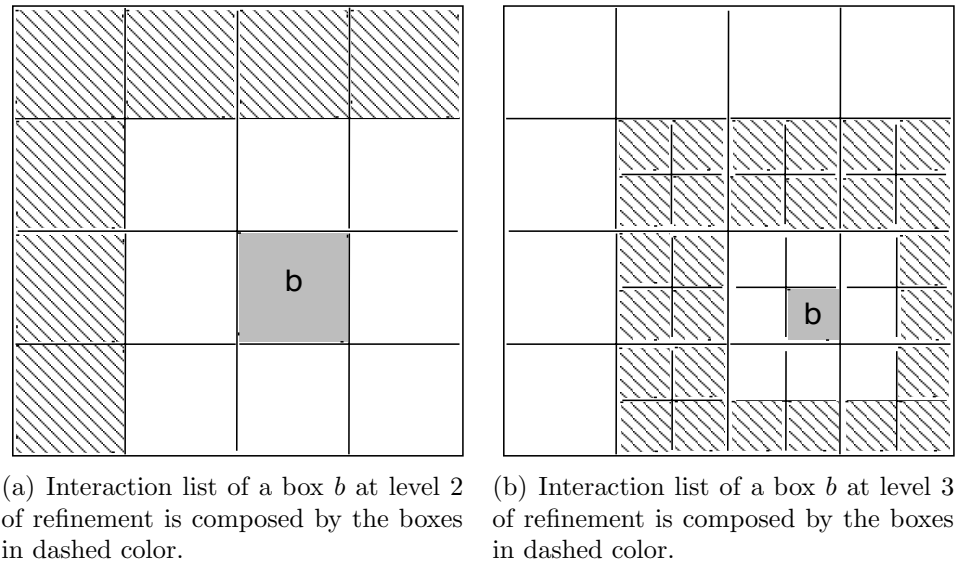

(a) Interaction list of a box $b$ at level 2 of refinement is composed by the boxes in dashed color.

(b) Interaction list of a box $b$ at level 3 of refinement is composed by the boxes in dashed color.

Figure 8: Interaction list for different levels: (a) Shows the interaction list of a box $b$ in a quad-tree refined until level 2. (b) Shows the further refinement of the boxes at the local list of the left figure, the members of the interaction list at level 3 are in dashed color, the box $b$ is in a solid gray color.

- *Child node.* A box $b$ at level $l$ is *child* of its *parent* box at level $(l-1)$.

- *Neighbor node.* The neighbors of a box $b$ at level $l$ are defined as any box that shares any boundary point with the box $b$ and that belongs to the same level than $b$. We can notice that there are at most 8 neighbors for any box at any level; only boxes situated on the borders of the computational domain have less than 8 neighbors.

- *Local list.* The local list of a box $b$ at a level $l$ is composed by the box $b$ itself and the boxes that are $b$'s neighbors.

- *Interaction list.* The interaction list of a box $b$ at level $l$ is composed by the boxes that are the children of the parent's neighbors, without including the boxes contained in the local list.

### 3.3.2 Translation of a Multipole Expansion (ME)

The FMM requires building an ME for every cluster of the system, and the clusters are the nodes from all the levels of the quadtree structure. The Multipole Epansions can be obtained directly just from the particles that compose the clusters,and the complexity to do so is $\mathcal{O}(lpN)$ and depends on the number of levels $l$, the truncation $p$, and the number of particles $N$.

But there is a more efficient way to obtain the ME for the clusters of the upper levels of the quadtree than directly building the ME from the particles. We can obtain the Multipole Expansion of a cluster $b$ by *translating* the ME of $b$'s children clusters, so the operation count for computing the ME for the cluster at a fixed level $l$ will depend on the cost of translating the ME of the cluster of the level beneath $(l+1)$.

The translation of the MEs is used to obtain the ME of the upper levels, levels $L-1$ to 2, in the hierarchy of the quad-tree. The ME of the parent's cell is obtained by shifting

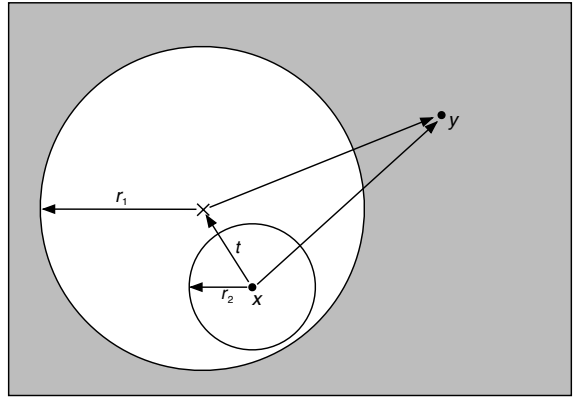

Figure 9: Translation of a Multipole Expansion.

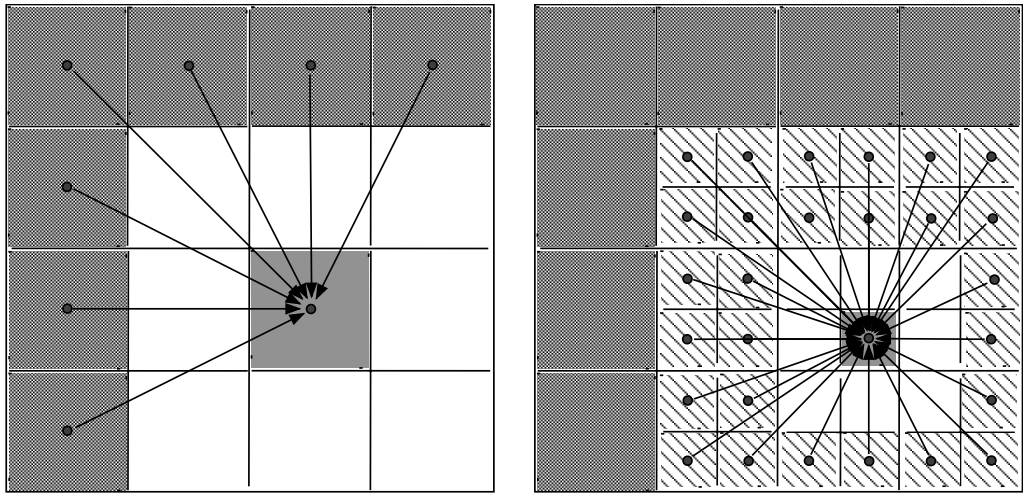

(a) Multipole Expansion to Local Expansion at Level 2

(b) Multipole Expansion to Local Expansion at Level 3

Figure 10: Interaction list for different levels: (a) Multipole Expansion to Local Expansion at Level 2 (b) Multipole Expansion to Local Expansion at Level 3.

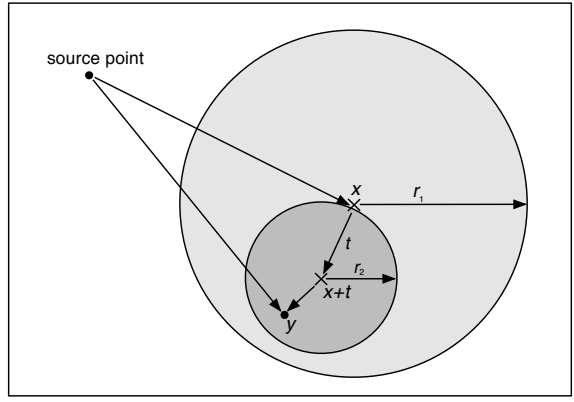

Figure 11: Translation of a Local Expansion.

the four children's ME to the center of their parent's cell. This procedure is called a "Multipole to Multipole" or M2M operation.

In matrix notation, the Multipole Expansion will resemble the following expression:

$$\Phi(y) = \sum_{m=0}^{p-1} A_m(x) Me_m(y) \tag{23}$$

By applying the translation operator we will have something like

$$\Phi(y) = \sum_{m=0}^{p-1} A_m(x) M2M(t) Me_m(y). \tag{24}$$

Where we have an expression with this form:

$$Me(y) = y^{-k-1} \tag{25}$$

that is shifted/translated by $t$:

$$\begin{aligned}
Me(y+t) &= (y+t)^{-k-1} = y^{-k-1}\left(1 + \frac{t}{y}\right)^{-k-1} \\
&= \sum_{n=k}^{\infty} \frac{(-1)^{n-k} n!}{k!(n-k)!} t^{n-k} Me(y) \\
&= \sum_{n=k}^{\infty} \binom{n}{k} Me(y)
\end{aligned}$$

Rewriting the equation in a vector matrix representation, and truncating the series, we obtain:

$$\begin{aligned}
Me(y+t) &= \sum_{n=k}^{p} \binom{n}{k} Me(y) \\
&= M2M(t) Me(y).
\end{aligned}$$

### 3.3.3 Translation of a Local Expansion

In the same way that we shift the Multipole Expansions of a child box to obtain the Multipole Expansion of the parent of that box, we can shift a Local Expansion of a parent box in order to add the influence of the far field of the parent box to the child box $b$. In this way we get a single representation of all the far domain for the box $b$.

By translating the LE of the upper level to the center of a box $b$, and adding the result to the LEs of the boxes in the interaction list of the same level than $b$, it is possible to obtain a single LE that represents the complete far field of the box $b$. This procedure can be applied to all the boxes of the quad-tree one level at a time, starting at level 2 and descending across the tree.

To add the parent's LE to a child LE it is necessary to shift the parent's LE to the center of its child. The procedure of translating the Local Expansions is called a "Local to Local" (L2L) operator.

A Local Expansion has this form:

$$Le_k(y) = y^k.$$

The idea is to translate the Local Expansion of a parent box to its children, as follows:

$$\begin{aligned} Le_k(y+t) &= (y+t)^k = \sum_{n=0}^{k} \frac{k!}{n!(k-n)!} t^{k-n} y^n \\ &= \sum_{n=0}^{k} \binom{k}{n} t^{k-n} Le_n(y) \end{aligned}$$

# 4 Sources of errors in the FMM as used in the vortex method

Before presenting our numerical experiments, let us review briefly the sources of error in the FMM approximation, as it applies to our 'client' application, the vortex particle method. First of all, we have approximated the Biot-Savart kernel by a $1/r$ kernel, as in [12]. This should not pose a problem, because the expansions in the FMM will be used at relatively large distances from the source points, and these kernels are a very good approximation to each other at these distances. We would like to keep this particular source of error as small as possible, at machine error in fact. To be able to accomplish this, as will be illustrated very clearly in some numerical experiments presented below, the size of the smallest boxes (at the deepest level of the tree) must maintain a minimum multiple of the particle characteristic sizes, $\sigma$.

## 4.1 Error by approximating the kernel

In order to obtain the velocity of $N$ vortex particles we need to calculate the following sum, using a kernel given below for the Gaussian particles:

$$u_\sigma(z) = \sum_{i=1}^{N} \Gamma_i \, \mathbf{K}_\sigma(z - z_i)$$

$$\mathbf{K}_\sigma(z) = \frac{j}{2\pi z^*} \left(1 - \exp\left(\frac{-|z|}{2\sigma^2}\right)\right) \tag{26}$$

Instead of using the original Gaussian kernel, we approximate it in the far field with the singular kernel:

$$u'(z) = \sum_{i=1}^{N} \Gamma_i \, \mathbf{K}'(z - z_i)$$

$$\mathbf{K}'(z) = \frac{j}{2\pi z^*} \tag{27}$$

The error introduced is:

$$\begin{aligned}
|u_\sigma(z) - u'(z)| &= \left| \sum_{i=1}^{N} \Gamma_i \, \mathbf{K}_\sigma(z - z_i) - \sum_{i=1}^{N} \Gamma_i \, \mathbf{K}'(z - z_i) \right| \\
&= \left| \sum_{i=1}^{N} \Gamma_i \, \left( \mathbf{K}_\sigma(z - z_i) - \mathbf{K}'(z - z_i) \right) \right| \\
&= \left| \sum_{i=1}^{N} \Gamma_i \left( \exp\left(\frac{-|z|}{2\sigma^2}\right) \right) \right|
\end{aligned} \tag{28}$$

If $a$ is the minimum distance between two particles for which the kernel is replaced by the singular $1/r$ kernel, an error bound can be computed:

$$|u_\sigma(z) - u'(z)| \leq \exp\left(\frac{-|a|}{2\sigma^2}\right) \left( \sum_{i=1}^{N} \Gamma_i \right) \tag{29}$$

When the distance between two particles is less than $a$, the original regularized Biot-Savart kernel is used.

## 4.2 Error by truncating the expansions

There is an approximation error related to the series expansion used to represent the kernel $1/r$ at large distances. The Taylor expansions are the most common choice; they are easier to deal with mathematically. But the Taylor series converges rather slowly, so many terms must be retained if one wishes to accomplish high accuracy (say, machine precision). This will have an impact on the efficiency of the method, as there is considerable computational overhead at the time of translating the expansions in both the upward and downward sweeps, when $p$ is large.

Finally, the error in the FMM approximation is dictated by the truncation level $p$.

The truncation at a certain coefficient $p$ introduces the error:

$$\left|[u(y)]^* - [u(y)]_p^*\right| = \left|[u(y)]^* - \sum_{m=0}^{p-1}\left(\sum_{i=1}^{N} -\frac{j\Gamma_i}{2\pi}(x_i - x_*)^m\right)(y - x_*)^{-m-1}\right|$$

$$= \left|\sum_{m=p}^{\infty}\left(\sum_{i=1}^{N} -\frac{j\Gamma_i}{2\pi}(x_i - x_*)^m\right)(y - x_*)^{-m-1}\right| \qquad (30)$$

Every source particle of the set $\{x_i\}$ is contained inside a circumference with center at $x_*$ and radius $r$, therefore for every particle of $\{x_i\}$ it is correct to say that $|x_i - x_*| < r$. We also know that for the ME (20) to converge, we must ensure that every evaluation point $\{y_j\}$ is located outside the circumference centered at $x_*$ and radius $r$. Thus, the source particles and the evaluation points must successfully fulfill the requirement that $|x_i - x_*| < |y - x_*|$. Furthermore, to obtain a good approximation of the initial kernel by using the expansion we should choose to use the ME only when the cluster with the source particles $\{x_i\}$ and the evaluation points $\{y_j\}$ meet a condition that ensures the ratio between the radius $r$ of the circumference $|x_i - x_*| < r$ and the distance from the center of the cluster to the evaluation point $|y_j - x_*|$ will be less than a fixed number $b$,

$$\left|\frac{r}{y_j - x_*}\right| \leq b. \qquad (31)$$

Then, for sets of source particles $\{x_i\}$ and evaluation points $\{y_j\}$ that comply with these conditions we can determine the theoretical error of using an ME to approximate the kernel,

$$\begin{aligned}
\left|[u(y)]^* - [u(y)]_p^*\right| &= \left|\sum_{m=p}^{\infty}\left(\sum_{i=1}^{N} -\frac{j\Gamma_i}{2\pi}(x_i - x_*)^m\right)(y - x_*)^{-m-1}\right| \\
&\leq \left|\sum_{i=1}^{N}\frac{\Gamma_i}{2\pi}\right|\left(\sum_{m=p}^{\infty}\left|\frac{r^m}{(y - x_*)^{m+1}}\right|\right) \\
&\leq A\left(\sum_{m=p}^{\infty}\left|\frac{r^m}{(y - x_*)^{m+1}}\right|\right) \\
&\leq A\left(\frac{1}{|y - x_*|}\sum_{m=p}^{\infty}\left|\frac{r^m}{(y - x_*)^m}\right|\right) \\
&\leq A\left(\frac{1}{|y - x_*|}\sum_{m=p}^{\infty}|b|\right) \\
&\leq A\left(\frac{1}{|y - x_*|}\frac{b^m}{1 - b}\right)
\end{aligned}$$

where,

$$A = \left|\sum_{i=1}^{N}\frac{\Gamma_i}{2\pi}\right|$$

We now have a formulation to calculate the *theoretical* maximum error related to the Multipole Expansion approximation of the kernel (17).

The effects of the different sources of error will be made evident in the numerical experiments presented in the next section.

# 5 Results of numerical experiments

In order to characterize the errors introduced by our approach, we present two different experimental setups, corresponding to more than 1000 experiments that were performed. For a number of illustrative computational experiments, we present the results in the form of the maximum observed errors and the spatial distribution of the errors, in order to reveal their relation with the algorithm parameters.

## 5.1 Lamb Oseen problem

The first problem setup corresponds to the use of the FMM in the context of an application of the vortex method to a viscous flow problem. We use the Lamb Oseen vortex, a known analytical solution of the Navier-Stokes equations, to initialize the amount of circulation $\Gamma_i$ of each of the particles of the system, as in [3]. The core sizes of the particles are uniform for all the particles and set to $\sigma = 0.02$. The particles of the system are positioned on a lattice distribution in a square domain of fixed size; the separation between particles is given by a constant spacing parameter $h$ and is obtained from the relation $\frac{h}{\sigma} = 0.8$, as in [3]. The analytical solution of the velocity field from the Lamb Oseen vortex is used to compare the results obtained with solving the Biot-Savart velocity, Equation (7), with our approach.

Equation (32) corresponds to the Lamb Oseen vortex solution in 2D:

$$\omega(r,t) = \frac{\Gamma_0}{4\pi\nu t} \exp\left(\frac{-r^2}{4\nu t}\right) \qquad (32)$$

where $r^2 = x^2 + y^2$. Figure 12 shows the spatial distribution of vorticity given by the Lamb Oseen vortex and the magnitude of the velocity field that it creates, with the parameters we have chosen, $\nu = 0.0005$ and $t = 4$.

We present the results obtained when using the Lamb Oseen vortex as the problem setup. The first set of graphs (Figures 13 and 14) corresponds to the maximum measured error of the FMM approximation, for different algorithm parameters and problem sizes. The FMM is used to compute the velocity at all $N$ particles of the system and the result is then compared against the analytical solution of the problem and normalized by the maximum velocity of the system. Each datapoint on the plots corresponds to a single experiment run for a given set of parameters. On these plots we can see the behavior of the measured error for the choice of level and truncation number in the algorithm, and the curves in the plots represent the change of the measured error at a fixed FMM level ($l$) when varying the truncation number ($p$). In total, for this experimental setup more than 900 experiments were performed and here we present the most representative results.

The set of experiments compares the results obtained for problems of different domain sizes. The increase in the domain size implies that more particles are needed to cover the space, as the particles are located on a lattice distribution with a fixed spacing $h$. The domain size affects the space partitioning and hierarchical structure of the FMM. Bigger domains require that the boxes of the quadtree be bigger in order to cover the domain. From the point of view of an evaluation point, a bigger domain implies that the near domain of a particle covers more space, and that the far field is located further away from the evaluation point.

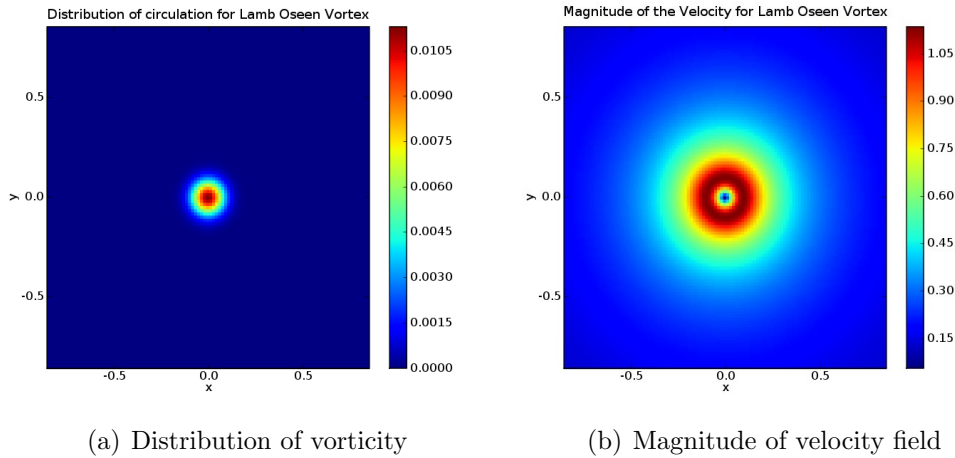

(a) Distribution of vorticity　　　　(b) Magnitude of velocity field

Figure 12: Distribution of vorticity and velocity magnitude given by the Lamb Oseen vortex solution.

Finally, the practical implication is that for bigger domains, the size of the most refined box of the FMM is wider, reducing the error of approximating the regular Biot-Savart kernel by $1/r$, as this error depends on the distance to the far field. This phenomena can be observed as the "flat" error lines on the plots for the higher levels $l$. The kernel error can be best observed on the more refined levels, where the far-field is closer.

An estimate to the error of approximating the kernel is given by Equation (29), which is proportional to the total amount of circulation of the particles in the system. Table 1 lists the maximum kernel error for a Lamb Oseen problem setup with 8836 particles. In order to build the estimate, the size of the box at the lower level of the queadtree structure is used as an estimate of the minimum distance between an evaluation point and the closest particle located in the far field that is going to be approximated.

Table 1: Approximation error estimate

| Total circulation | FMM level | FMM box size | Approximation error (log10) |
|---|---|---|---|
| 1 | 2 | 0.375 | -76.768 |
| 1 | 3 | 0.187 | -19.211 |
| 1 | 4 | 0.093 | -4.596 |
| 1 | 5 | 0.046 | -0.717 |

Figures 15 , 16, 17, and 18 characterize the spatial structure of the error of the FMM approximation for a Lamb Oseen problem setup with 11449 particles. Each plot corresponds to a single experiment with fixed parameters (level of refinement $l$ and truncation number $p$), the magnitude of the $log10$ error obtained when evaluating the velocity for every particle is represented with a color. The plots of the spatial distribution of the error reveal the spatial structure of the approximations and the close relationship of the error with the data structure used in the algorithm. On the plots we can see the error structure that depends on the approximation of the far field and the one that depends of

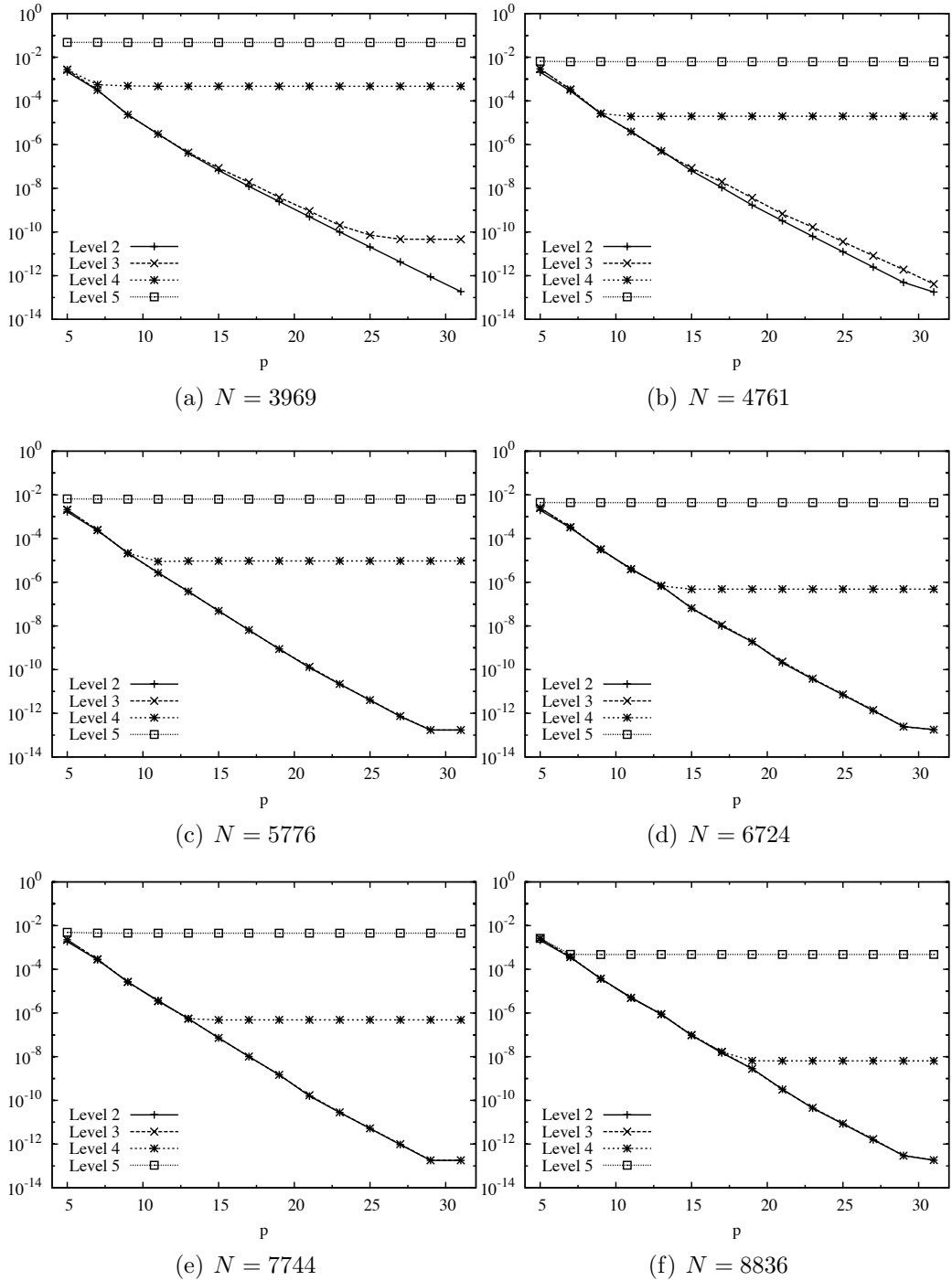

Figure 13: Maximum measured error of the FMM calculation of the velocity, with respect to the analytical value at every evaluation point. Each marker represents one full evaluation of the FMM velocity, with a given choice of level $l$ and truncation $p$ on the abscissa.

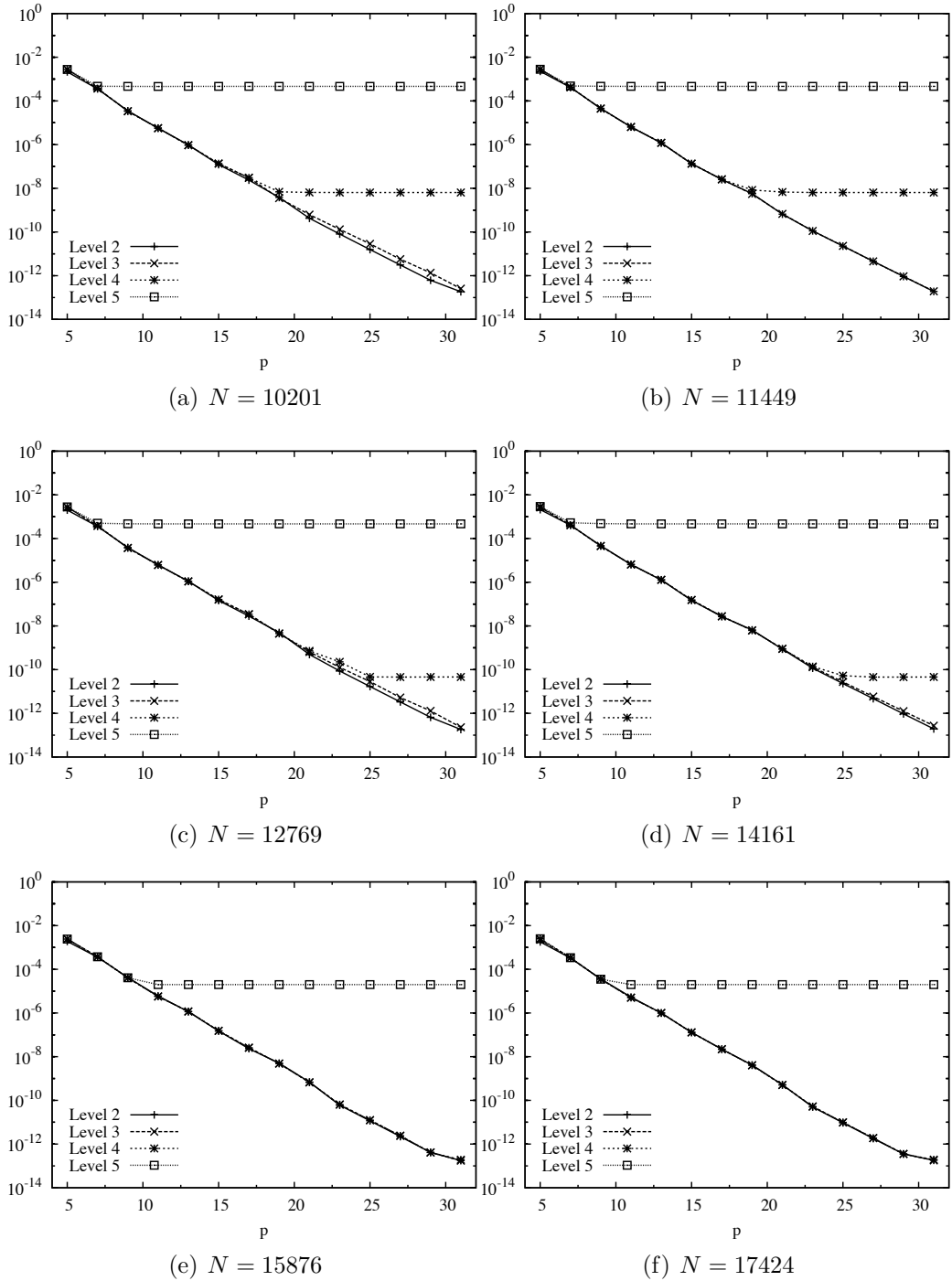

Figure 14: Maximum measured error of the FMM calculation, with respect to the analytical value of the velocity at every evaluation point. Each marker represents one full evaluation of the FMM velocity, with a given choice of level $l$ and truncation $p$ on the abscissa.

the numbers on terms retained for the Multipole Expansion ($p$).

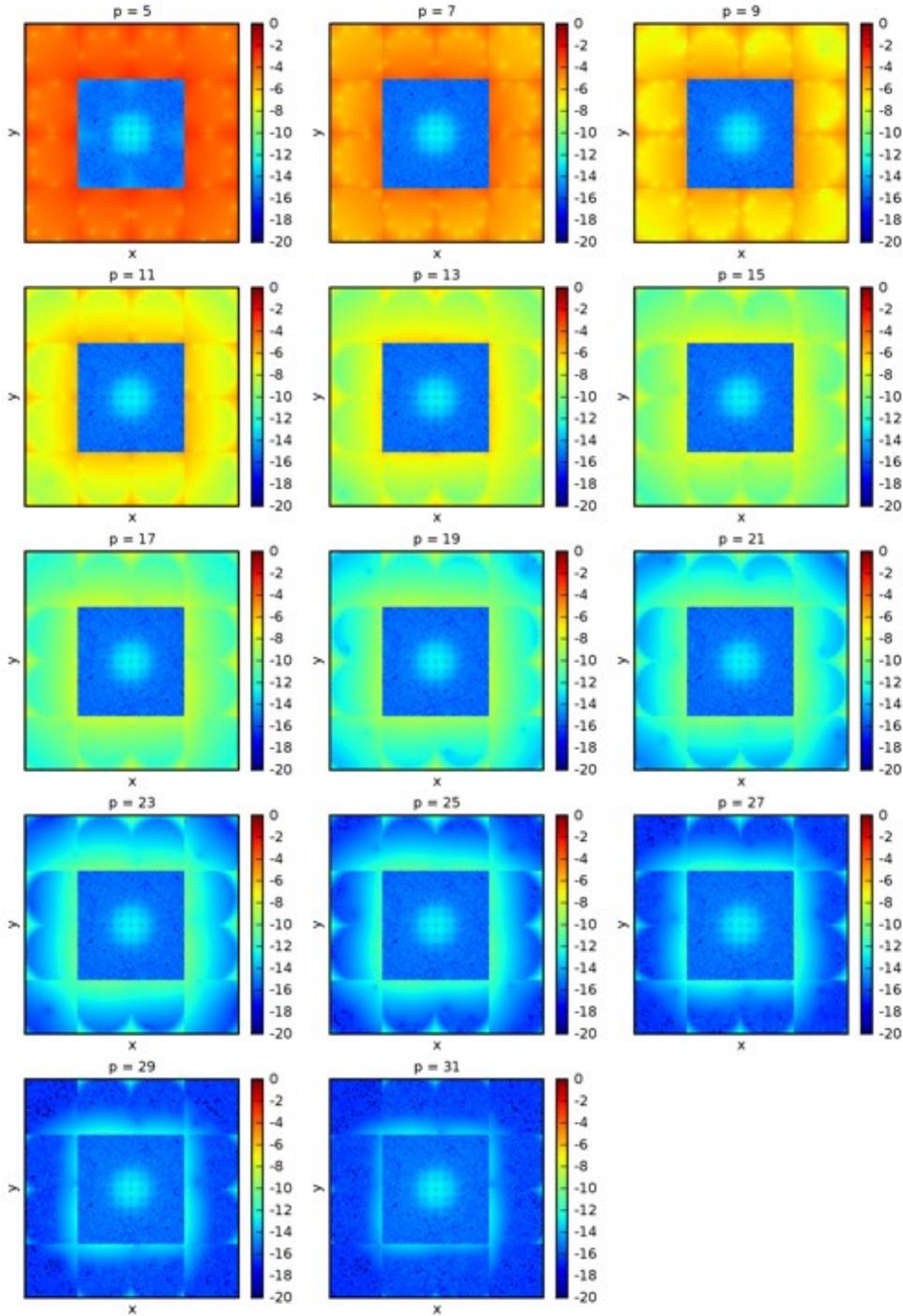

Figure 15: Evolution of the spatial distribution of the error made by the FMM when compared against the analytical solution of the test problem Lamb Oseen with problem size 11449, FMM hierarchy up to level 2 and varying the number $p$ of Multipole Coefficients retained.

To focus on the change of the error of approximating the kernel with its singular version when the distance to the far field increases, we compare the Lamb Oseen setup at three experiments with the same parameters for level of refinement ($l$) and truncation number ($p$) but different problem sizes: 3969, 7744, and 19044 particles. Using the setup described before, figure 19 presents more clearly the effects of the approximation of the far field using the singular kernel, and as the the size of the domain increases the distance to the far field also grows, decreasing the effect of the kernel error.

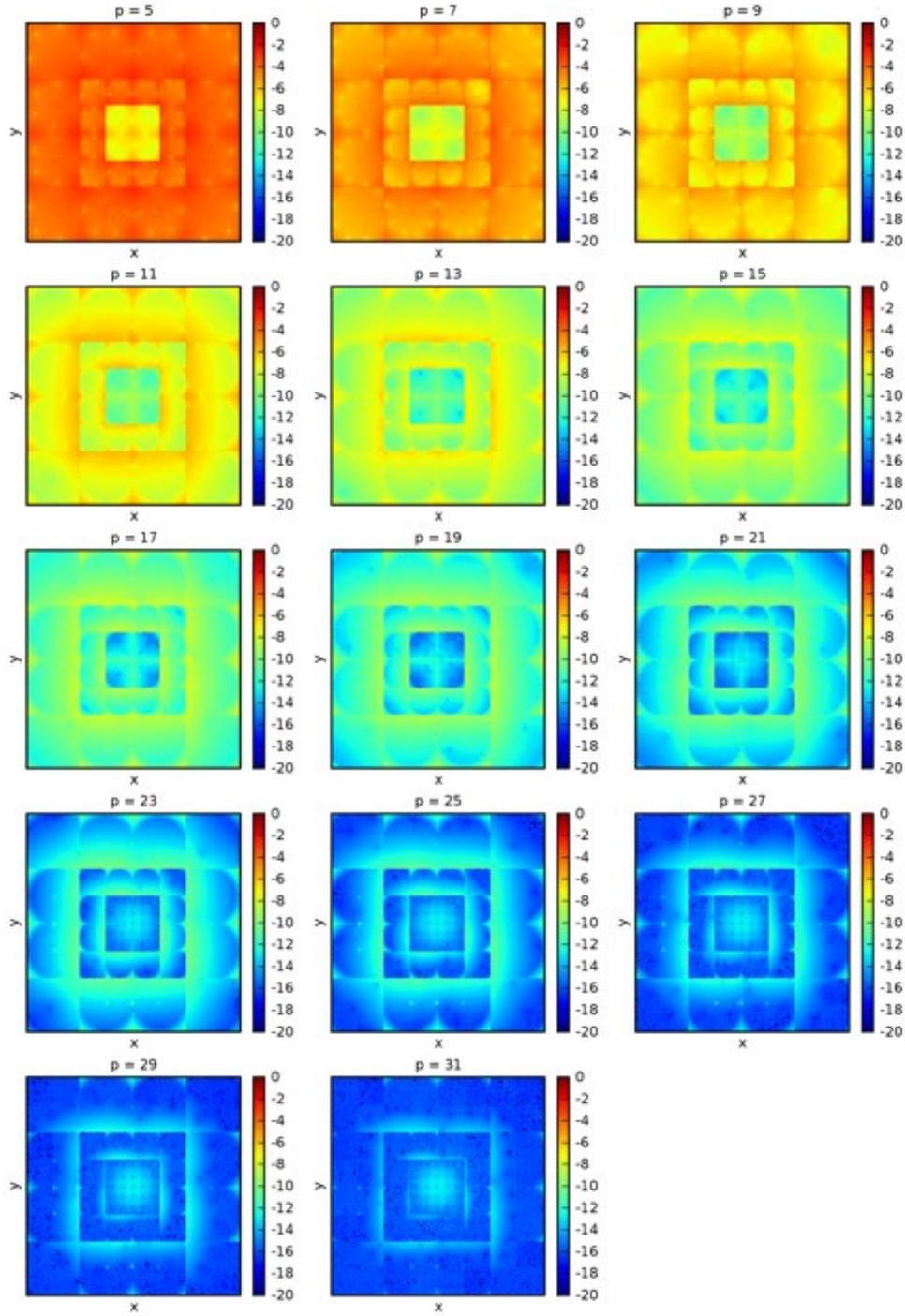

Figure 16: Evolution of the spatial distribution of the error made by the FMM when compared against the analytical solution of the test problem Lamb Oseen with problem size 11449, FMM hierarchy up to level 3 and varying the number $p$ of Multipole Coefficients retained.

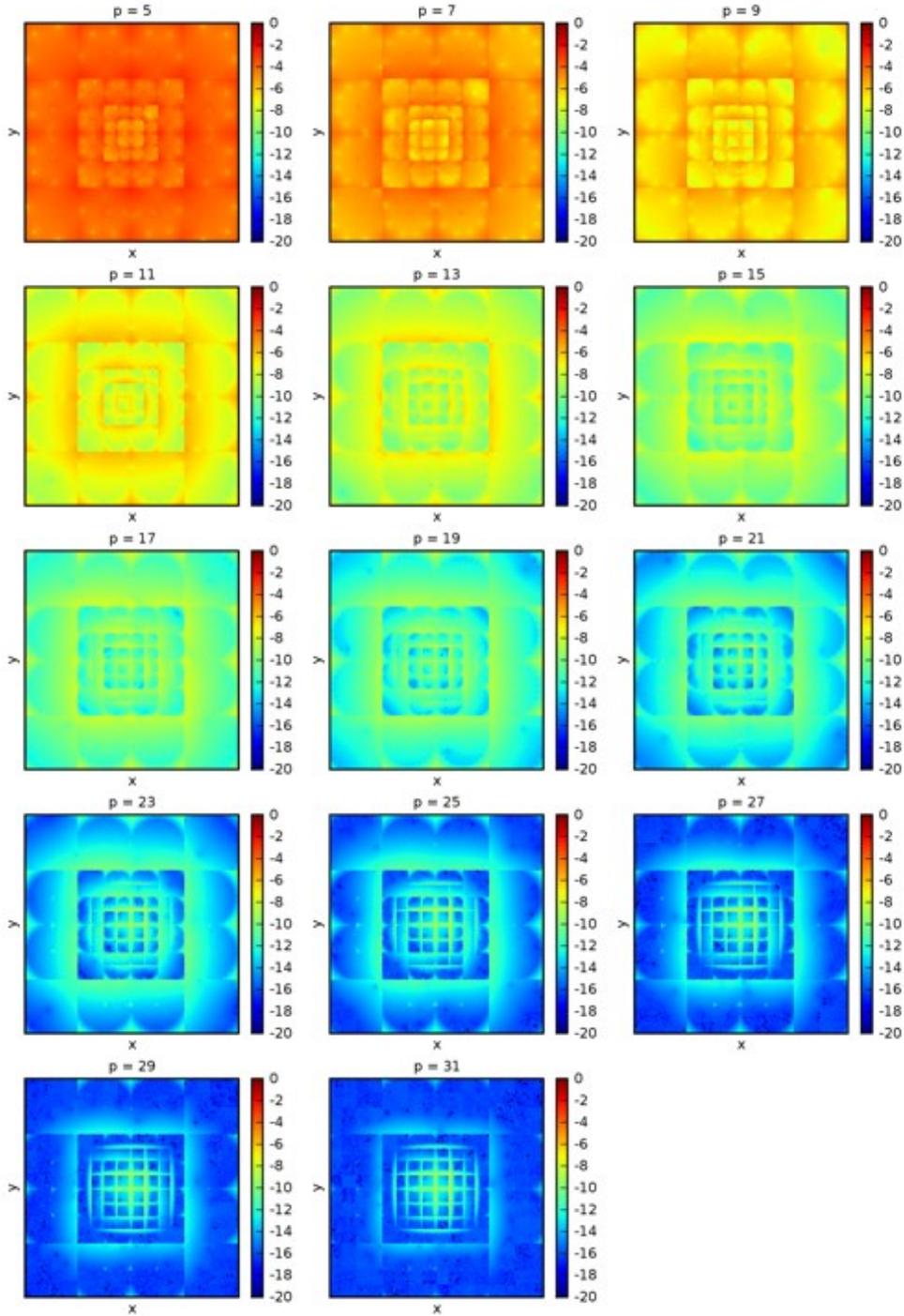

Figure 17: Evolution of the spatial distribution of the error made by the FMM when compared against the analytical solution of the test problem Lamb Oseen with problem size 11449, FMM hierarchy up to level 4 and varying the number $p$ of Multipole Coefficients retained.

## 5.2 Single source problem

The second problem setup corresponds to a single source particle that has nonzero weight and is located at the center of the domain. Evaluation points are located on a lattice distribution across a square domain, and the result obtained from the FMM calculation is compared against the direct solution of the particles interaction. We do this in order to evaluate the approximation obtained by means of the FMM in the domain.

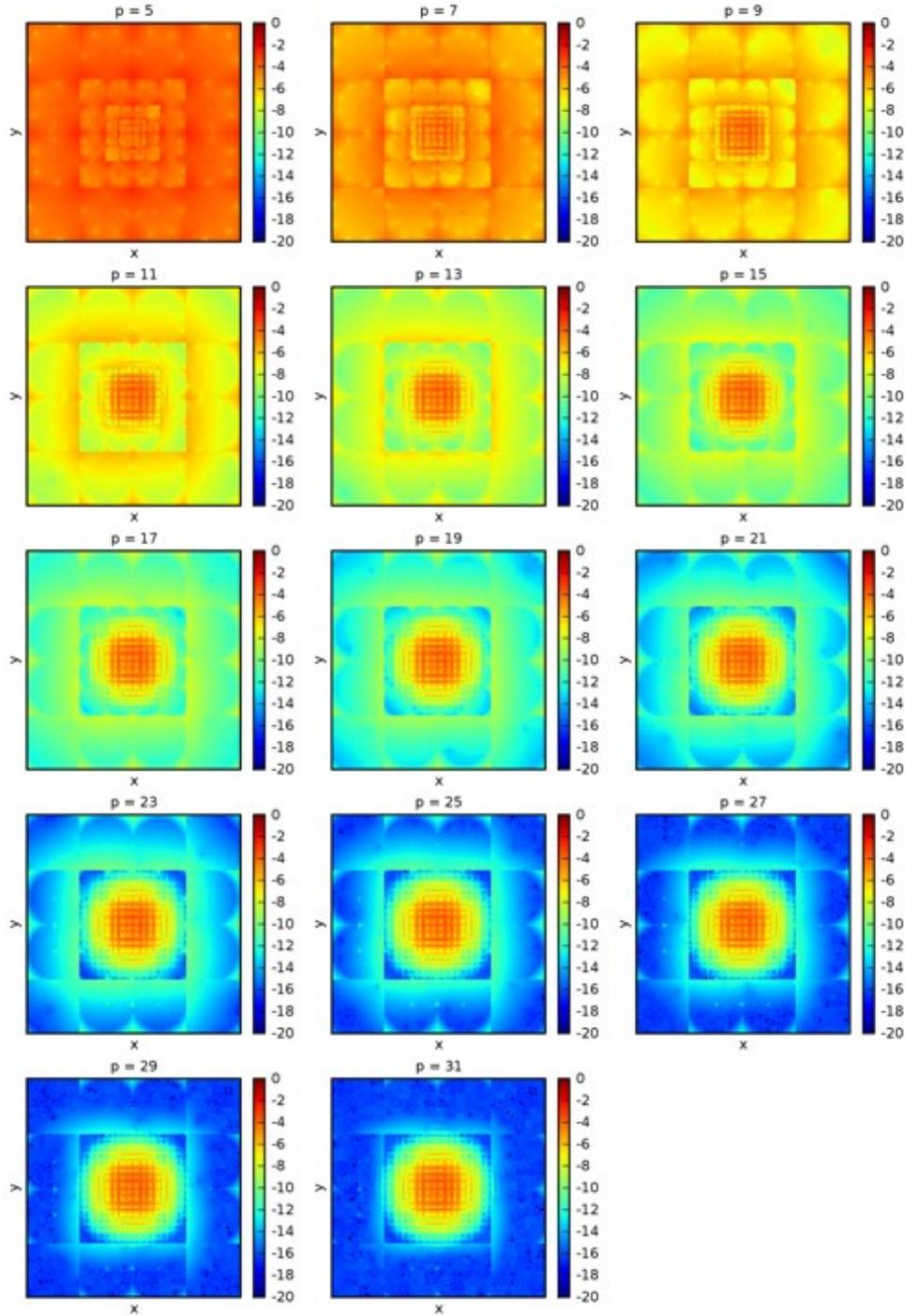

Figure 18: Evolution of the spatial distribution of the error made by the FMM when compared against the analytical solution of the test problem Lamb Oseen with problem size 11449, FMM hierarchy up to level 5 and varying the number $p$ of Multipole Coefficients retained.

Figures 20, 21, 22 show the change of the spatial distribution of the error incurred by the FMM for a problem with a single source particle. On each figure, we compare the

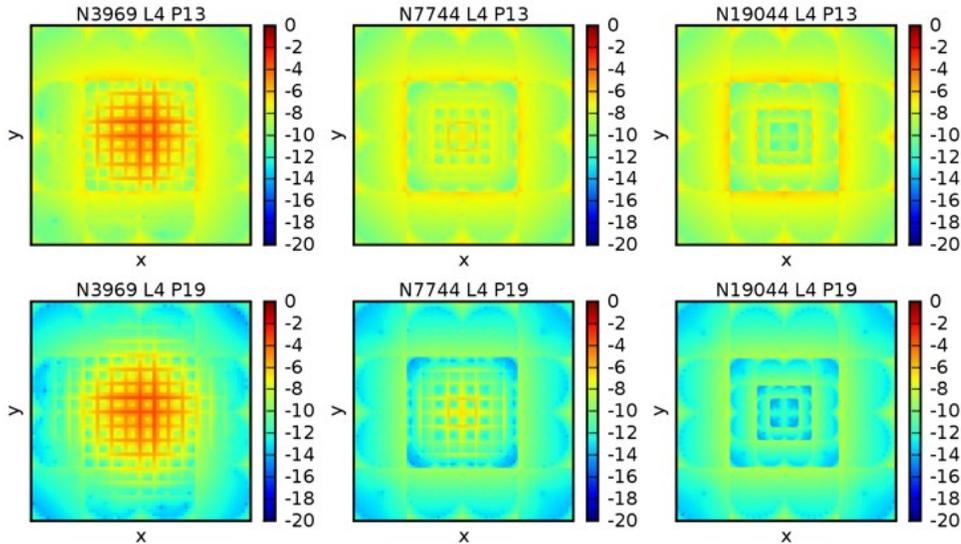

Figure 19: Evolution of the distribution of the error when the distance of the far domain approximation increases at a given level $l$ and truncation point $p$. Columns show experiments for problem sizes $N = (3969, 7744, 19044)$ from left to right. First row, has parameters $l = 4$ and $p = 13$. Second row, has parameters $l = 4$ and $p = 19$.

distribution of the error when more terms of the multipole expansion are kept for a fixed level, where each plot presents a single experiment for a fixed set of parameters (level and truncation point). Figures 20, 21, 22 present the experiments for 8836 evaluation points for levels 3, 4 and 5 respectively. On any given plot, on the one hand we have that the error for the near domain is machine error, as the near domain is computed directly and no FMM error is introduced, and on the other hand for the far field we can see the error behavior introduced by the Multipole Expansion and Local expansion constructions.

The kernel error is easily observed as a *high error* circumference near the center of the plot, it is clearly visible for high values of $p$ in Figure 22, though the error can also be observed in Figure 21. The kernel error dominates when the far-field is not located *far* enough, and as we increase the distance to where the far field is considered this error decreases. Figure 23 illustrates the same problem for a fixed truncation number $p$ and different levels, exposing the relation between the level of refinement of the FMM and the kernel error more explicitly.

# 6  Conclusions

The Fast Multipole Method (FMM) has offered dramatic increase in the computation capability for all processes dominated by pair-wise interactions, or $N$-body problems. In many applications, such as gravitational systems, high accuracy may not be an issue, and furthermore such a system is dominated strongly by the first moment (due to the fact that all mass is positive). In other applications, the accuracy of the overall numerical scheme may have been dominated by factors such as spatial discretization or time stepping errors. However, as high-order methods become more commonplace, and as the availability of powerful computers becomes more widespread, scientists may have the need to know and control the accuracy of the FMM approximation more finely.

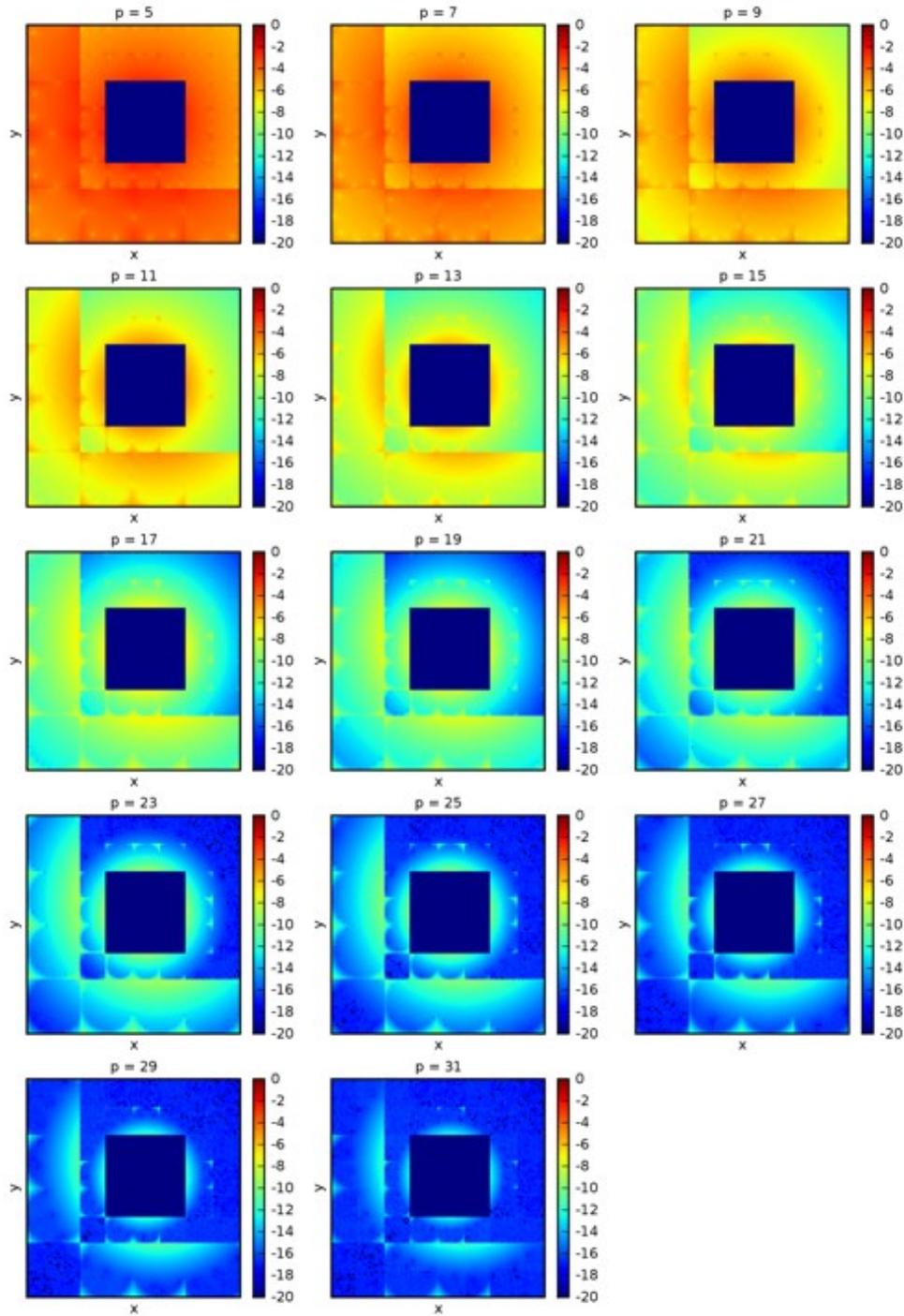

Figure 20: Evolution of the spatial distribution of the error made by the FMM for a problem with a single particle as a source and 8836 evaluation points distributed in a lattice, FMM hierarchy up to level 3 and varying the number $p$ of Multipole Coefficients retained.

The literature on the FMM algorithm supplies theoretical error bounds, however these may not be very tight and moreover they are not very illustrative as to the behavior of the method, except for providing simple ideas like decrease of error with increasing truncation number. In the literature, one rarely finds reports of *measured* errors when using the algorithm. This of course is not practical in most cases, as a comparison with either an analytical solution or with the direct $\mathcal{O}(N^2)$ summation would be required. We take precisely this approach, and use some simple experimental setups to discover the behavior of the FMM errors with respect to the different parameters and choices that the user would normally make: number of particles, levels in the tree, truncation level.

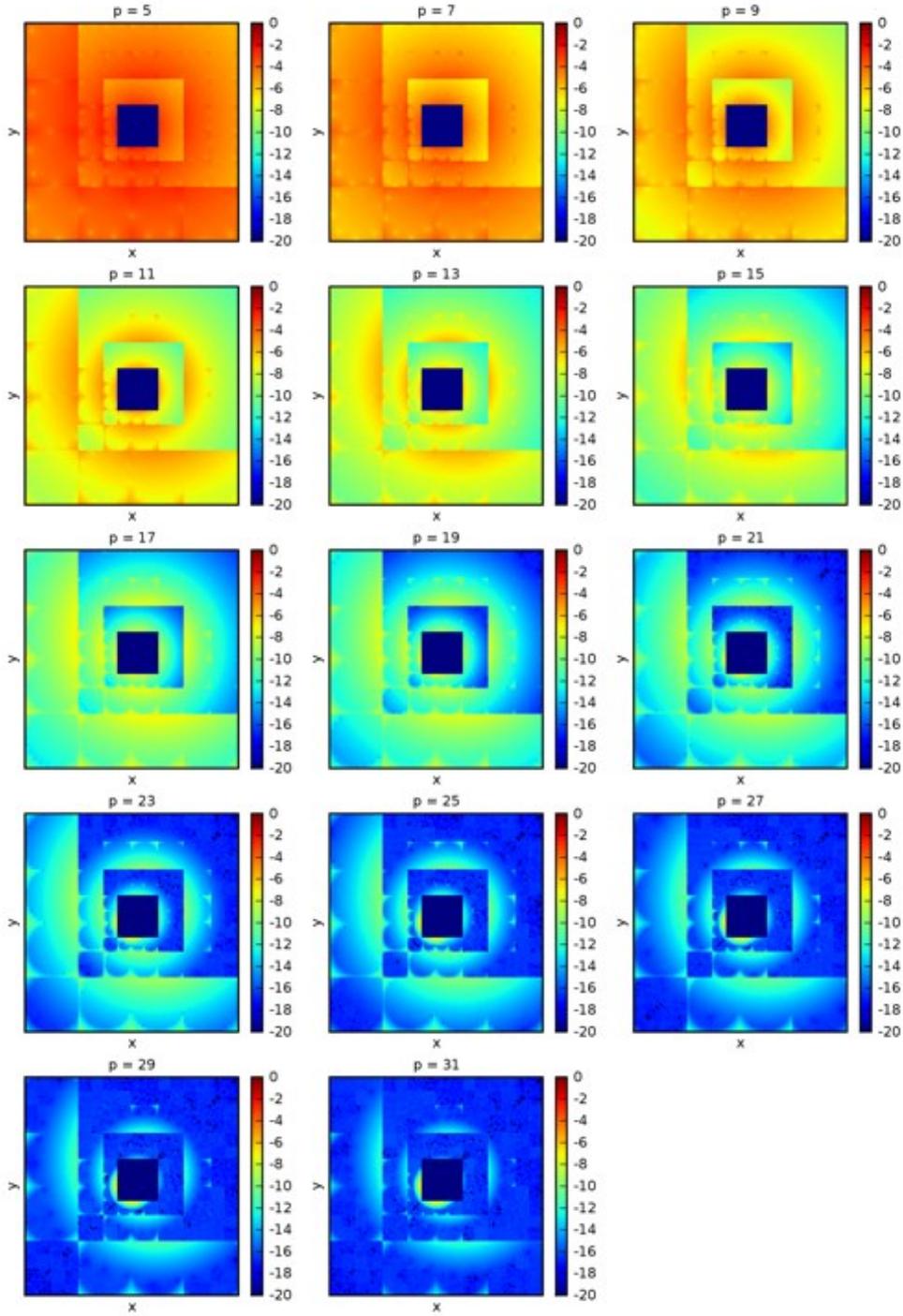

Figure 21: Evolution of the spatial distribution of the error made by the FMM for a problem with a single particle as a source and 8836 evaluation points distributed in a lattice, FMM hierarchy up to level 4 and varying the number $p$ of Multipole Coefficients retained.

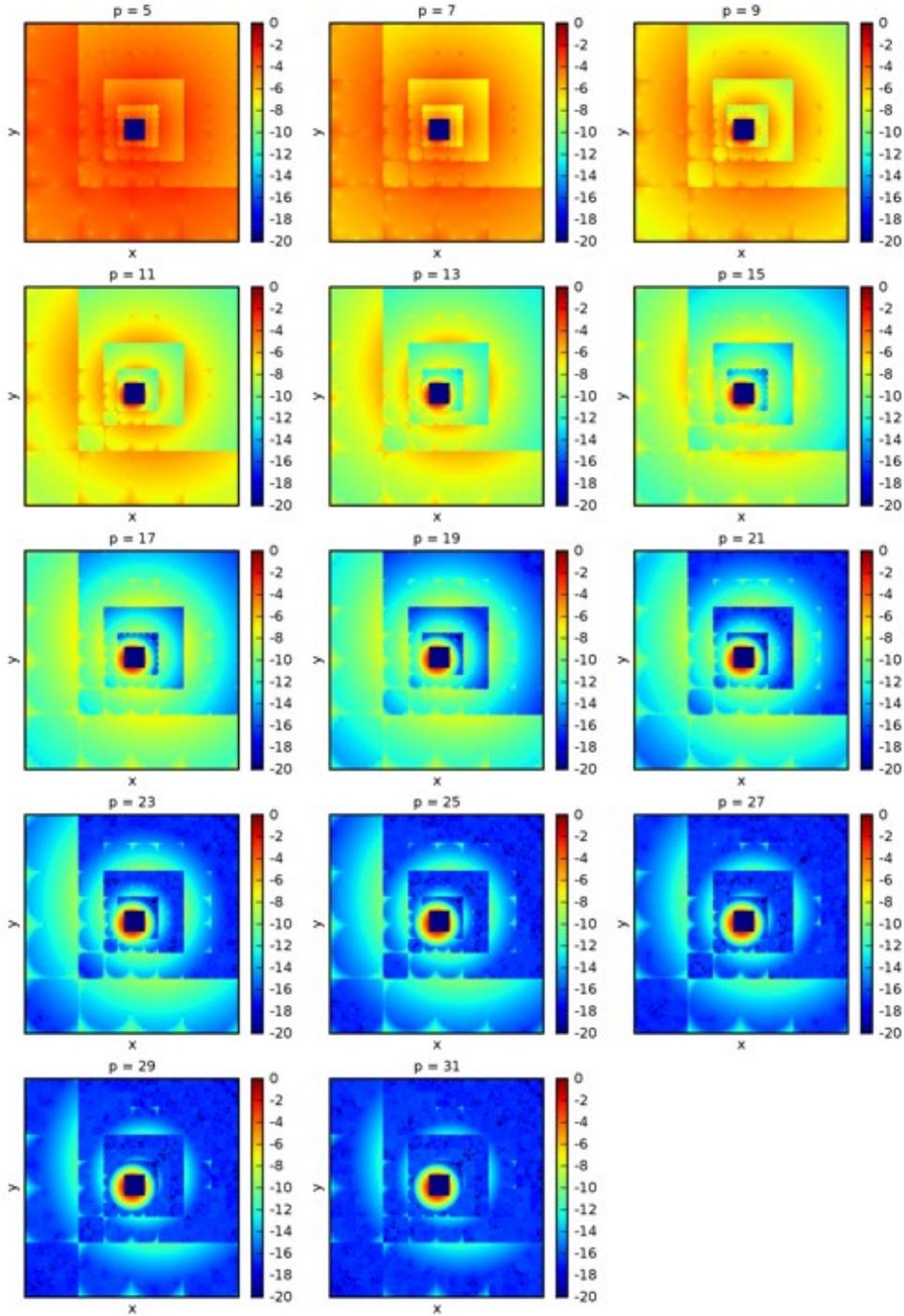

Figure 22: Evolution of the spatial distribution of the error made by the FMM for a problem with a single particle as a source and 8836 evaluation points distributed in a lattice, FMM hierarchy up to level 5 and varying the number $p$ of Multipole Coefficients retained.

Users of the FMM will have a *client* application which will provide them with a given kernel. One can approximate the kernel by another (such as $\frac{1}{r}$) with a good accuracy, but this implies a need to keep box sizes at the deepest level at a size large compared with the kernel width. The kernel change has the aim of simplyfying the mathematics of the expansions. It makes sense to reduce the human effort required for the implementation, in this way, but only if the consequences of the approximation are well understood by the practitioner. We believe that the experiments presented in this work illustrate nicely the result of approximating the force or influence kernel by a simpler one which applies in the far field.

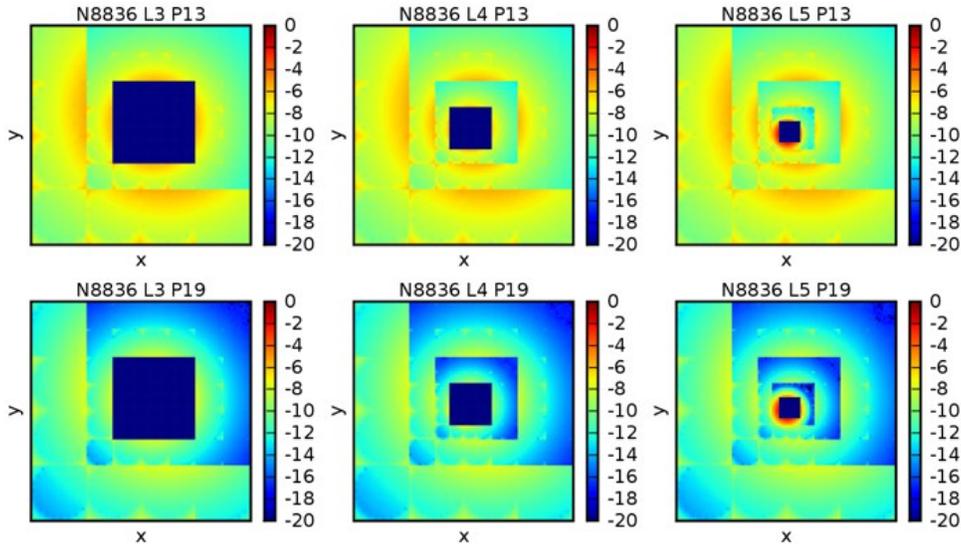

Figure 23: Evolution of the distribution of the error when the refinement increases for a problem of fixed size. In the Image is possible to see the evolution of the error as we vary the number of levels used by the FMM, columns from left to right show levels 3, 4 and 5 respectively. Row one and two compares the error against the FMM truncation numbers 13 and 19 respectively.

The spatial distribution of the error incurred by the FMM offers visual evidence of all the contributing factors to the overall approximation accuracy: multipole expansion, local expansion, hierarchical spatial decomposition (interaction lists, local domain, far domain). This presentation is a contribution to any researcher wishing to incorporate the FMM acceleration to their application code, as it aids in understanding where accuracy is gained or compromised.

Potential users of the FMM may wish to perform some experiments similar to those presented here, varying algorithm parameters, to get a good grasp of the error behavior for their problems. In this sense, we offer an experimental design which will help users apply the FMM algorithm with good knowledge of what it is doing.

In addition to the understanding of the accuracy of the FMM approximation, a user would also require knowledge of the efficiency obtained, as of course there is always a trade-off between accuracy and computational efficiency. We have not presented speed-up tests or measures of the calculation times, because using a Python prototype which is not optimal in terms of speed may offer misleading results. Moreover, the problem sizes which we manage are small and likely not in the range of noticeable acceleration with the FMM. Realistically, one will observe the expected speed-up of the algorithm with tens or hundreds of thousands of particles. In this range of problem size, experimentation aiming to measure the errors of the algorithm would be very cumbersome (unless an analytical solution were available).

In conclusion, we suggest that whenever the application calls for controlled and high accuracy, the user may want to follow the example of this work to get to characterize the errors for his/her application, and then follow this by some speed up tests at larger problem sizes.

We are making our Python FMM code available (via Google Code) to the community, and welcome any correspondence with interested readers.

To end, we offer an opinion with regards to the areas of future progress in the field. The FMM algorithm is mature and optimal, but still rather difficult to program. We anticipate that there remain the following areas for progress:

1. Algorithm acceleration through hardware. There are already a few researchers investigating, for example, the use of graphical processing units (GPUs) with the FMM. We are also initiating some work in this area.

2. Better representation of kernels, that provide more accuracy with less truncation number. As already mentioned, the Taylor series are the easiest to deal with mathematically, but converge rather slowly. In applications demanding high accuracy, perhaps other series representations would provide a better solution, as long as the human effort required for the mathematical derivations and programming are acceptable.

3. Generality and portability. We refer to having tools that work with different kernels and the availability of software library components for wider dissemination and impact of the algorithm in multiple applications. In this context, we have initiated collaboration with the development team of the PETSc library to produce a parallel and portable version of the FMM which can be distributed to the wider community of computational scientists.